\documentclass[11pt,a4paper]{article}
\usepackage{jheppub}
\usepackage{epsfig,multicol,bbm,amsxtra,subfigure,ulem}
\title{{New localization method of $U(1)$ gauge vector field on flat branes in (asymptotic) $AdS_{5}$ spacetime}}

\author[a]{Zhen-Hua Zhao\footnote{Corresponding author},}
\author[b,c]{Qun-Ying Xie,}
\author[c,d]{Yuan Zhong}
\affiliation[a]{
    Department of Applied Physics,
    Shandong University of Science and Technology,
    Qingdao, 266590 People's Republic of China}
\affiliation[b]{School of Information Science and Engineering,
Lanzhou University, Lanzhou 730000, People's Republic of China }
\affiliation[c]{Institute of Theoretical Physics,
        Lanzhou University, Lanzhou 730000,
        People's Republic of China
    }
\affiliation[d]{IFAE, Universitat Aut\`onoma de Barcelona, 08193 Bellaterra, Barcelona, Spain}
\emailAdd{zhaozhh09@lzu.edu.cn}
\emailAdd{xieqy@lzu.edu.cn}
\emailAdd{yzhong@ifae.es}

\abstract{It is well known that the $U(1)$ gauge vector field, with the standard five-dimensional (5D) action, cannot be localized on  Randall-Sundrum-like braneworlds with an infinite extra dimension. In this paper, we propose a modified 5D action to localize $U(1)$ gauge vector field on flat branes with an infinite or finite extra dimension. The localization method is realized by adding a dynamical mass term into the standard 5D
action of the vector field, which is proportional to the 5D scalar curvature. It is shown that the vector zero mode is localizable if the 5D spacetime is (asymptotic) $AdS_{5}$. Moreover, the  massive tachyonic modes can be excluded.
}


\begin{document}
\maketitle
%\flushbottom

\section{Introduction}

The braneworld theories have received a lot of attentions since the success of the {Randall}-Sundrum (RS) thin brane models \cite{Randall199983,Garriga2000}.
In braneworld theories, the localization of {gravity and} all kinds of
matter fields {on the brane} is always an important issue. It is well known that the {four-dimensional} graviton can be localized in
the RS thin brane scenario  \cite{Randall199983,Randall199983a} and in thick
brane scenarios \cite{Gremm2000478,Bazeia200405,Barbosa-Cendejas200510,Barbosa-Cendejas200673,FarakosKoutsoumbasPasipoularides2007,GermanHerrera-AguilarMalagon-MorejonMora-LunaRocha2012,Herrera-AguilarMalagon-MorejonMora-LunaQuiros2012}.
The localization of real scalar field {is the same as gravity in general relativity~\cite{Bajc2000}, but may be different from or even opposite to gravity in some modified gravity} \cite{ZhongLiuYang2010, YangLiuZhongDuWei2012}.
By introducing the usual Yukawa coupling between the background scalar field and the fermion field, the fermion can also be localized on{the brane generated by an odd scalar field} \cite{Rubakov1983,Bajc2000,Randjbar-Daemi2000,Ringeval200265,Koley200522,Melfo2006,LiuZhangZhangDuan2008}.
{However, if the brane is generated by an even scalar field, we need to introduce new localization mechanism in order to localize fermion on the brane}\cite{LiuXuChenWei2013}.

% when the background scalar field has a kink-like configuration interpolating between the two different vacua
%at the two sides of brane.
%However, when the scalar field is an even function of the extra dimension, one need to introduce the new localization mechanism presented in ref.~\cite{LiuXuChenWei2013}.

In five-dimensional (5D) spacetime, the $U(1)$ gauge field {$\mathcal{A}_{M}$} with the usual action
\begin{equation}
S\sim\int d^5x\sqrt{-g}\mathcal{F}_{MN}\mathcal{F}^{MN}, \label{action0}
\end{equation}
where $\mathcal{F}_{MN}=\partial_M \mathcal{A}_N-\partial_N \mathcal{A}_M$ is the field strength, can be localized {on the brane} in some special braneworld models, for example, in the standing wave braneworld model \cite{GogberashviliMidodashviliMidodashvili2012},
 in the  braneworld models with finite {extra dimension} \cite{Liu200808,Liu20090902,LiuFuGuoLi2012, GuoHerrera-AguilarLiuMalagon-MorejonMora-Luna2013,Herrera-AguilarRojasSantos-Rodriguez2014}, and in a 6D model \cite{Oda2000}.
But it cannot be localized in a RS-like baneworld model with an infinite {extra dimension}  \cite{Pomarol2000,Bajc2000,ChumbesHoffHott2012}.

In order to localize $U(1)$ gauge field on branes in 5D {RS-like} models with {an infinite extra dimension}, the typical {localization mechanism}  is to reform the action \eqref{action0}. In the thin brane {scenario}, many ideas {were} proposed {for} this issue \cite{Oda2001,Ghoroku2002,Giovannini2002,Guerrero2009}. In ref. \cite{Oda2001}, the author {added} a topological term and a 3-form gauge potential into the action \eqref{action0}. In ref. \cite{Ghoroku2002}, a bulk mass term of the vector and a
coupling between the vector potential and the brane {were introduced}. In ref. \cite{Giovannini2002},  the action \eqref{action0} {was} changed to
\begin{equation}
S\sim\int d^5x\sqrt{-g}e^{2\alpha(y)}\mathcal{F}_{MN}\mathcal{F}^{MN},\label{action1}
\end{equation}
where $e^{2\alpha(y)}$ is the warp factor. In this model the {vector} zero mode can be localized on the negative tension brane. In ref. \cite{Guerrero2009},  gauge field kinetic terms induced by localized fermions {were} added {into} the gauge field action.

In order to localize $U(1)$ vector fields in a thick brane models, Kehagias and Tamvakis (KT) {proposed} a coupling between the gauge field and an extra dilaton field. The KT mechanism has been applied in many different
brane-world scenarios
\cite{CruzTahimAlmeida2010,AlencarLandimTahimMunizCosta2010,CruzLimaAlmeida2013,FuLiuGuo2011,CruzTahimAlmeida2009,ChristiansenCunhaTahim2010,CruzMalufAlmeida2013}.
Recently, Chumbes, Holf da Silva and Hott (CHH) {proposed} a new mechanism \cite{ChumbesHoffHott2012}, in which gauge and tensor fields directly couple to a functional of the background scalar field.  By introducing a Stueckelberg-like action, Vaquera-Araujo and Corradini realized the localization of vector in thick brane model\cite{Vaquera-AraujoCorradini2014}.

%On the other hand, the thick brane is usually generated by a background scalar field. In ref. \cite{Bazeia200405}, Bazeia and Gomes introduced the Bloch brane generated by two real scalar fields. This brane model was further generalized in ref. \cite{SouzaDutra200878}, and investigated in refs.~\cite{Gomes2006, CorreaDutraHott2010,  CruzLimaAlmeida2013, CruzMalufAlmeida2013,  XieYangZhao2013,  LiuXuChenWei2013}. It is known that $U(1)$ gauge field with the action (\ref{action0}) can not be localized on the Bloch brane \cite{CruzLimaAlmeida2013}. In ref. \cite{CruzLimaAlmeida2013} the localization of guage field on the Bloch brane was discussed with the KT mechanism. In order to localize the zero mode on the Bloch brane, an extra dilaton scalar field is introduced in ref. \cite{CruzLimaAlmeida2013}.

In this paper, we propose a new {localization method}  to localize $U(1)$ gauge field on the brane.
%In our {localization method}  a 5D  dynamic mass term of gauge field is
%introduced in the 5D action of gauge field. The 5D  dynamic mass term is directly proportional
%to the 5D scalar curvature.
{We assume the 5D gauge field has a dynamic mass term, which is proportional to the 5D scalar curvature.}
Our {localization method}  can be used both in thin and thick braneworld models with infinite or finite {extra dimension}. To localize gauge field zero mode, the only assumption {we need} is that the
5D spacetime is (asymptotic) $AdS_{5}$.
%In our method the tachyon modes are excluded, and this is free to the above assumption.
{With the same assumption, we can further prove that there is no tachyonic mode in the KK spectrum.}

%. The appearance of that mass term in the action will destroy the $U(1)$ gauge symmetry, but which will not destroy the same symmetry of the 4D effective action for the zero mode gauge field, and it holds 5D Lorenz invariance.

The paper is structured as follows. In {section} \ref{sec-2}, we show the setup of our {localization method of gauge field on the brane} and prove that the zero mode is localizable {under the assumption of that} the {five-dimensional spacetime} is (asymptotic) $AdS_5$. Further, in section \ref{tachyon}, we prove that { tachyon modes} can be excluded. {We conclude our results in the last section.}

\section{{Localization method of vector field on the branes}}
\label{sec-2}

{The} line-element {describing a flat (Minkowski) braneworld embedded in} five-dimensional spacetime is assumed to be {\cite{Randall199983}}
\begin{eqnarray}
ds^2=\mathcal{G}_{MN}dx^Mdx^N=e^{2\alpha(y)}\eta_{\mu\nu}dx^{\mu}dx^{\nu}+dy^2,\label{ymetric}
\end{eqnarray}
where $e^{2 \alpha(y)}$ is the warp factor, $\alpha(y)$ is {a} function of {the extra dimension} $y$, $\mathcal{G}_{MN}$ is the metric of 5D bulk {spacetime} {and $M, N=0,1,2,3,4$ stand for the bulk coordinate indices}, and the Minkowski metric $\eta_{\mu\nu}$ on branes with signature $(-1,+1,+1,+1)$ and $\mu, \nu=0,1,2,3$ correspond to the brane coordinate indices.

 In order to localize vector fields on branes,  {we} introduce a dynamical mass term {and the action reads}
 \begin{equation}
 S=\int d^4x dy\sqrt{-\mathcal{G}}\left (-\frac{1}{4}\mathcal{G}^{MN}\mathcal{G}^{RS}\mathcal{F}_{MR}\mathcal{F}_{NS}
 -\frac{1}{2}\mathcal{M}^2\mathcal{G}^{MN}\mathcal{A}_M \mathcal{A}_N\right ),\label{vectorAction}
 \end{equation}
where $\mathcal{M}^2$ is a function of the 5D scalar curvature.
%The appearance of that mass term in above action will destroy the $U(1)$ gauge symmetry, but which will not destroy the same symmetry of   4D effective action for the zero mode gauge field.
Here, we set
\begin{equation}
\mathcal{M}^{2}=-\frac{1}{16}\mathcal{R}, \label{M2}
\end{equation}
 {With} the metric in Eq. \eqref{ymetric}, {the scalar curvature is given by}
\begin{equation}
\mathcal{R}=-4(5\alpha'^{2}+2\alpha'').\label{5Ry}
\end{equation}
{In this paper, the prime} stands for the derivative with respect to $y$.
Because the brane is flat so $\mathcal{R}$ is only a function of extra-dimension coordinate $y$.
%From the above action (\ref{vectorAction}) one can get the equation of motion  {for the vector field} $\mathcal{A}_M$,
%\begin{equation}
%\frac{1}{\sqrt{-\mathcal{G}}}\partial_{M}\left(\sqrt{-\mathcal{G}}\mathcal{G}^{MN}\mathcal{G}^{RS}\mathcal{F}_{NS}\right)-\mathcal{M}^2\mathcal{G}^{MR}\mathcal{A}_M=0.
%\end{equation}

Here, we parametrize  $\mathcal{A}_{M}$ in the following way \cite{BatellGherghetta2006,Ghoroku2002,AlencarLandimTahimCosta2014}:
\begin{equation}
\mathcal{A}_{M}=(\hat{\mathcal{A}}_{\mu}+\partial_{\mu}\phi,\;\mathcal{A}_{4}), \label{eqAM}
\end{equation}
where $\hat{\mathcal{A}}_{\mu}$ is the transverse component which satisfies the transverse condition $\partial_{\mu}\hat{\mathcal{A}}^{\mu}=0$, and $\phi$ is the longitudinal component. Substituting Eq. \eqref{eqAM} into the action \eqref{vectorAction}, with the transverse condition, one can find that the transverse vector $\hat{\mathcal{A}}_{\mu}$ decouples from the scalar fields $\phi$ and $\mathcal{A}_{4}$, and the action \eqref{vectorAction} can be split into two parts,
 \begin{equation}
  S=S_{V}(\hat{\mathcal{A}}_{\mu})+S_{S}(\hat{\mathcal{A}}_4,\,\phi),
 \end{equation}
where \cite{Peeters2007a,Peeters2007}
\begin{eqnarray}
S_{V}=\int d^4x dy\sqrt{-\mathcal{G}} \left(- \frac{1}{4}\, \hat{\mathcal{F}}_{\lambda \mu} \hat{\mathcal{F}}_{\nu \rho} {\mathcal{G}}^{\lambda \nu} {\mathcal{G}}^{\mu \rho} - \frac{1}{2} {\partial}_{4}{\hat{\mathcal{A}}_{\mu}}  {\partial}_{4}{\hat{\mathcal{A}}_{\nu}} {\mathcal{G}}^{4 4} {\mathcal{G}}^{\mu \nu} - \frac{1}{2} \mathcal{M}^{2} \hat{\mathcal{A}}_{\mu} \hat{\mathcal{A}}_{\nu} {\mathcal{G}}^{\mu \nu}\right),\label{actionT}
\end{eqnarray}
\begin{eqnarray}
S_{S}=\int d^4x dy\sqrt{-\mathcal{G}} \Big(&-& \frac{1}{2} {\partial}_{4}({\partial}_{\mu}{\phi})  {\partial}_{4}({\partial}_{\nu}{\phi})  {\mathcal{G}}^{4 4} {\mathcal{G}}^{\mu \nu} - \frac{1}{2} \mathcal{M}^{2}  {\partial}_{\mu}{\phi}  {\partial}_{\nu}{\phi}  {\mathcal{G}}^{\mu \nu} 
- \frac{1}{2} {\partial}_{\mu}{{\mathcal{A}}_{4}}  {\partial}_{\nu}{{\mathcal{A}}_{4}}  {\mathcal{G}}^{4 4} {\mathcal{G}}^{\mu \nu} \nonumber\\
&-& \frac{1}{2} {\mathcal{A}}_{4} {\mathcal{A}}_{4} \mathcal{M}^{2} {\mathcal{G}}^{4 4}
 + {\partial}_{\mu}{{\mathcal{A}}_{4}}  {\partial}_{4} ({\partial}_{\nu}{\phi})  {\mathcal{G}}^{4 4} {\mathcal{G}}^{\mu \nu}\Big),
\end{eqnarray}
and $\hat{\mathcal{F}}_{\mu\nu}=\partial_{\mu}\hat{\mathcal{A}}_{\nu}-\partial_{\nu}\hat{\mathcal{A}}_{\mu}$.
%One can further decompose $\mathcal{\mathcal{A}}_{4}$ as 
%\begin{equation}
%\mathcal{A}_{4}=\psi+\partial_{4}\phi,
%\end{equation}

%\begin{eqnarray}
% S&=&\int d^4x dy\sqrt{-\mathcal{G}} \left(- \frac{1}{4}\, {F}_{\mu \nu} {F}_{\lambda \rho} {g}^{\mu \lambda} {g}^{\nu \rho} - \frac{1}{2}\, {\partial}_{4}{\hat{A}\,_{\mu}}\,  {\partial}_{4}{\hat{A}\,_{\nu}}\,  {g}^{4 4} {g}^{\mu \nu} - \frac{1}{2}\, M \hat{A}\,_{\mu} \hat{A}\,_{\nu} {g}^{\mu \nu}\right)\nonumber\\
% & +&\int d^4x dy\sqrt{-\mathcal{G}}\Big( - \frac{1}{2}\, {\partial}_{\mu}{\psi}\,  {\partial}_{\nu}{\psi}\,  {g}^{4 4} {g}^{\mu \nu}  - \frac{1}{2}\, M {\partial}_{\mu}{\phi}\,  {\partial}_{\nu}{\phi}\,  {g}^{\mu \nu} - \frac{1}{2}\, M \psi \psi {g}^{4 4} \nonumber \\
% & & -M {\partial}_{4}{\phi}\,  \psi {g}^{4 4} - \frac{1}{2}\, M {\partial}_{4}{\phi}\,  {\partial}_{4}{\phi}\,  {g}^{4 4}\Big),\label{vectorAction2}
%\end{eqnarray}

Because we focus on the localization of vector field, and the scalar sector had been discussed in Ref. \cite{AlencarLandimTahimCosta2014}, therefore we will not discuss the localization of the scalar particles  in this paper.
The transverse vector field can be decomposed as 
\begin{equation}
\hat{\mathcal{A}}_{\mu}(x,y)=\sum_{n}A^{(n)}_{\mu}(x)\chi_{n}(y),\label{KKdeco}
\end{equation}
where  $A^{(n)}_{\mu}(x)$ is the 4D vector KK mode and $\chi_{n}(x)$ is called as the KK wave function \cite{Ponton2012}, here and the following ``$\Sigma_{n}$'' is a shorthand for the summation and integration over discrete and continuum KK modes.

By means of the KK decomposition \eqref{KKdeco}, the action \eqref{actionT}  can be further reduced to %{the effective action of a massless and a series of massive four-dimensional vector fields:}
\begin{eqnarray}
S_{V}&=&-\frac{1}{4}\sum_{n}
     \int dy \chi_n^2\int d^4x
     \big({\eta^{\mu\nu}\eta^{\lambda\rho}
          F^{(n)}_{\mu\lambda} F^{(n)}_{\nu\rho}
          +}2 m_n^2 {\eta^{\mu\nu}A^{(n)}_{\mu} A^{(n)}_{\nu}}
     \big)\label{action5t},
\end{eqnarray}
where $F^{(n)}_{\mu\nu}=\partial_{\mu}A^{(n)}_{\nu}-\partial_{\nu}A^{(n)}_{\mu}$ is the four-dimensional vector field strength tensor. In order to obtain the action \eqref{action5t}, the KK wave function $\chi_n(y)$ is required to  satisfy the following equation
\begin{equation}
-\partial_y\left(e^{2\alpha}\partial_y \chi_n\right)+\chi_n e^{2\alpha}\mathcal{M}^2=m_n^2\chi_n, \label{eq}
\end{equation}
and the orthonormalization condition
\begin{equation}
\int {\chi}_m(y) {\chi}_n(y)  dy = 0, \quad(m\neq n). %\delta_{mn}.
\end{equation}
From the actions \eqref{action5t}, the localization of the 4D vector KK mode {requires}
\begin{equation}
I\equiv\int_{-\infty}^{+\infty}  \chi_n^2(y) dy<\infty \label{int}.
\end{equation}
{By using} the following {field} transformation
\begin{equation}
\chi_n=e^{-\alpha}\tilde{\chi}_n,
\end{equation}
Equation. \eqref{eq} can be {rewritten as}
\begin{eqnarray}
-\tilde{\chi}''_n+\big(\alpha''+\alpha'^2+ \mathcal{M}^2-e^{-2\alpha}m_n^2\big)\tilde{\chi}_n=0.\label{eq2}
\end{eqnarray}

For the {vector} zero mode, $m_{0}=0$, Eq. \eqref{eq2} can be written as
\begin{equation}
-\tilde{\chi}''_0+\big(\alpha''+\alpha'^2+ \mathcal{M}^2\big)\tilde{\chi}_0=0. \label{eq4}
\end{equation}
Substituting Eqs. \eqref{M2} and \eqref{5Ry} into { Eq. \eqref{eq4}}, we can obtain
\begin{equation}
-\tilde{\chi}''_0+\big(\frac{3\alpha''}{2}+\frac{9\alpha'^2}{4}\big)\tilde{\chi}_0=0, \label{eq5}
\end{equation}
 which can be further factorized as
\begin{eqnarray}
\bigg(-\frac{d}{dy}-\frac{3}{2}\alpha'\bigg)\bigg(\frac{d}{dy}-\frac{3}{2}\alpha'\bigg)\tilde{\chi}_0=0. \label{eq3}
\end{eqnarray}
The solution of the above equation is
\begin{equation}
\tilde{\chi}_0(y)={c_0}e^{\frac{3}{2}\alpha(y)},
\end{equation}
and further one can get
\begin{equation}
\chi_0=e^{-\alpha}\tilde{\chi}_0={c_0}e^{\frac{1}{2}\alpha}. \label{zeromode1}
\end{equation}
With the above solution, the integration \eqref{int} reads
\begin{equation}
I=\int_{-\infty}^{+\infty}  \chi_0^2dy={c_0^2}\int_{-\infty}^{+\infty} e^{\alpha}dy \label{int31}.
\end{equation}
Because the integrand in Eq. \eqref{int31}
is continuous at the interval $y\in(-\infty,+\infty)$, the convergency of the above integrations is determined by the asymptotic behaviors of integrands at the infinity.

{Now, we prove} that if the {five-dimensional spacetime} is asymptotic $AdS_5$, the vector zero mode is {always} localizable. {The condition of asymptotic $AdS_5$ means that}, when $y\rightarrow\pm\infty$, {the warp factor has the following asymptotic behavior}
\begin{eqnarray}
&&\alpha(y)|_{y\rightarrow\pm \infty}\rightarrow -k|y|,\label{alphas}
\end{eqnarray}
where {$k$ is a positive constant and it usually relates with the five-dimensional fundamental scale in RS-like braneworld scenarios}. With this condition, the asymptotic {behavior} of integrand in Eq. \eqref{int31} reads as
\begin{equation}
e^{\alpha}|_{y\rightarrow\pm \infty}\rightarrow e^{-k|y|}.
\end{equation}
%\textcolor{blue}{Therefore} we have the limitation
%\begin{equation}
%\lim\limits_{\pm\infty}y^p f(y)=0,
%\end{equation}
%where $p=2>1$.
Therefore, the integration \eqref{int31} is convergent, namely, the 4D vector zero mode is localizable. If the {extra dimension} is finite, the conclusion is also valid.

Our conclusion can be easily verified in the RS-II model \cite{Randall199983a}, in which $\alpha(y)=-k|y|$. Substituting it into {Eq. \eqref{int31}}, we obtain
{\begin{eqnarray}
I% &=&\int_{-\infty}^{+\infty}  \chi_n^2(y) dy\nonumber\\
 = 2 c_0^2 \int_{0}^{+\infty}  e^{-ky} dy
    %+\int_{-\infty}^{0}  e^{ky} dy
 =   2c_0^2/{k}. \label{int3}
\end{eqnarray}
Normalizing $I$ results in $c_0=\sqrt{k/2}$.}
Therefore, the vector zero mode is localizable in RS-II model.

%Above all, for a brane-world model with the five-dimensional spacetime is asymptotic $AdS_5$, the zero mode of $U(1)$ guage field is localizable.

%For the massive KK modes, we cannot obtain the analytic form of $\chi_n$ from equation \eqref{eq2}.
%The massive solution of $\chi_n$ can be obtained by the numerical method with exact braneworld models in following sections.

\section{Excluding 4D vector tachyonic modes}\label{tachyon}

Further, we will discuss if this new localization { method}  can exclude {the tachyonic modes} of 4D vector field.
To  discuss the {massive modes and mass spectrum}, it is more convenient to work with the conformal metric:
\begin{eqnarray}
  ds^2=e^{2\alpha(z)}\left(\eta_{\mu\nu}dx^\mu dx^\nu+dz^2\right).\label{conmetric}
\end{eqnarray}
%this is just a  coordinate transformations: $dz\equiv e^{-\alpha}dy$ and $z=\int_{0}^{y}e^{-\alpha(y)}dy$.
The form of dynamic mass with above metric \eqref{conmetric} is
\begin{equation}
\mathcal{M}^2=-\frac{1}{16}\mathcal{R}=\frac{1}{4}e^{-2\alpha}(3\dot\alpha^{2}+2\ddot\alpha). \label{Mz}
\end{equation}
where the symbol `` $\spdot$ " means the derivative {with} respect to $z$.

Making use of the KK decomposition
\begin{equation}
\hat{\mathcal{A}}_{\mu}(x,z)=\sum_{n}A^{(n)}_{\mu}(x)\chi_{n}(z),
\end{equation}
and the transformation
\begin{equation}
\chi_{n}(z)={\rho}_{n}(z)e^{-\alpha/2},
\end{equation}
the action \eqref{actionT} reduces to
\begin{equation}
S_{V}=-\frac{1}{4}\sum_{n}\int dz \rho_n(z)^2\int d^4x\big(F^{(n)}_{\mu\nu}F^{(n),\mu\nu}+2 m_n^2 A^{(n)}_{\mu}A^{(n),\mu}\big)\label{action6}.
\end{equation}
The localization of KK modes means
\begin{equation}
I\equiv\int_{-\infty}^{+\infty}  \rho_n^2(y) dy<\infty \label{int8}.
\end{equation}
At the same time, {the} function $\rho(z)$ is required to satisfy  a Schr\"odinger-like equation
\begin{equation}
\left(-\partial^2_{z}+V(z)\right)\rho_{n}(z)=m^2_{n}\rho_{n}(z)  \label{Erho}
\end{equation}
and the orthonormalization condition
\begin{equation}
\int {\rho}_m(z) {\rho}_n(z)  dz = 0 ~~~(m\neq n). %\delta_{mn}.
\end{equation}
 {The effective potential is given by},
\begin{equation}
V(z)=\frac{1}{2} \ddot{\alpha}(z)+\frac{1}{4}\dot\alpha(z)^2+ e^{2\alpha}\mathcal{M}^{2}. \label{Vz}
\end{equation}
Substituting {the expression of $\mathcal{M}^{2}$ in \eqref{Mz} into Eq. (\ref{Erho})}, we obtain
\begin{equation}
V(z)= \ddot{\alpha}(z)+\dot\alpha(z)^2. \label{Vz2}
\end{equation}
With above potential {Eq.} (\ref{Erho}) can be further factored as
\begin{equation}
\left(-\partial _z-\dot\alpha\right)  \left(\partial _z -\dot\alpha \right)\rho_{n}(z)=m^2_{n}\rho_{n}(z).\label{eqmz}
\end{equation}
This is the supersymmetry quantum mechanics form of the Schr\"odinger equation (\ref{Erho})  \cite{Andrianov2003},  which guarantees the positivity of $m_n^2$.
Therefore, the tachyonic vector modes are excluded. {This result is independent of the asymptotic behavior of spacetime.}
%And this \textcolor{blue}{needless to require that the 5D spacetime is (or asymptotic) $AdS_{5}$}.
%\sout{\textcolor{blue}{is model independent.}}

\section{Conclusion}
In this paper,
we {proposed} a new {localization method}  of $U(1)$ gauge field. In our { method}  a dynamical mass term {was} added into the {5D action} of the vector field.
{The} dynamical mass term is {proportional}
to the 5D scalar curvature.
{It was shown that, if the brane is embedded in a 5D $AdS_{5}$ spacetime, then} the {vector} zero mode {is localized on the brane}. Moreover, {we also proved that there is no vector tachyonic mode with our method.}

%A mass term appearing in action means the $U(1)$ gauge symmetry is destroyed in 5D spacetime,
%but this will not destroy the $U(1)$ gauge symmetry of the 4D effective action of the zero mass vector.

\section*{Acknowledgments}

We thank Prof. Y.-X. Liu for helpful discussion. Z.-H. Zhao is supported by the National Natural Science Foundation of China (Grant No.  11305095), the Natural Science Foundation of Shandong Province, China (Grant No. ZR2013AQ016), and Scientific Research Foundation of Shandong University of Science and Technology for Recruited Talents (Grant No. 2013RCJJ026).
Q.Y. Xie is supported by the National Natural Science Foundation of China (Grant No. 11375075).
Y. Zhong is supported by the scholarship granted by the Chinese Scholarship Council (CSC).

\bibliographystyle{JHEP}
%\bibliographystyle{apsrev}
%\bibliographystyle{apsrmp4-1}
%\bibliography{E:/360yun/library_zhao/articles_all}
%\bibliography{/Users/zhaozhenhua/yunpan/library_zhao/articles_all}
\bibliography{VecLoc}

\end{document}